# Mode-resolved logarithmic quasiballistic heat transport in thin silicon layers: Semianalytic and Boltzmann transport analysis

Jae Sik Jin*

Department of Smart Manufacturing Systems
Chosun College of Science & Technology
Gwangju 61453, Republic of Korea

*Author to whom correspondence should be addressed.
*E-mail address*: jinjs@cst.ac.kr.

## ABSTRACT

Nonequilibrium phonon transport driven by nanoscale hotspot heating in silicon device layers governs heat dissipation in advanced microelectronics and underscores the need for a better microscopic understanding of such processes. Yet the origin of the frequently observed logarithmic (ln) dependence of the apparent thermal response on hotspot size in crystalline silicon, and the role of individual phonon modes in this regime, remain unclear. Here, we develop a semianalytical, mode-resolved framework in the spectral phonon mean free path (MFP) domain and validate it against a full-phonon-dispersion Boltzmann transport model for heat removal from a $10 \times 10\ \mathrm{nm}^2$ hotspot in a thin Si layer (thicknesses of 41, 78, and 177 nm) representative of a silicon-on-insulator transistor. We show that ln-type quasiballistic scaling arises only for modes that lie on a log-uniform conductivity plateau and are diffusive-side or quasiballistic with respect to the hotspot size, whereas fully ballistic long-MFP modes contribute a saturated, nonlogarithmic background, leading to extremely slow suppression of their heat-carrying capability. The resulting phonon-modal nonlocal spectrum establishes spectral selection rules for ln-regime transport in confined Si and provides a compact basis for incorporating mode-selective quasiballistic corrections into continuum thermal models and for interpreting phonon-resolved thermometry experiments.

## I. INTRODUCTION

When the characteristic length scale of a temperature or power variation becomes comparable to or smaller than the phonon mean free paths (MFPs), heat transport enters a nonlocal regime in which Fourier's law with a constant thermal conductivity is no longer adequate. In this regime, the effective thermal conductivity depends on a wave vector $k$ or on a characteristic length $L_c$, and solutions of the phonon Boltzmann transport equation (BTE) have shown that logarithmic ("ln-scale") behavior can arise. For example, in Debye-type models with realistic scattering [e.g., phonon relaxation time $\tau \propto \omega^{-3}$, where $\omega$ is the phonon angular frequency], the nonlocal thermal conductivity $K(k)$ can exhibit a logarithmic dependence $K(k) \propto \ln(1/k)$,[1] and thin-film or low-dimensional systems can show $K(L_c) \propto \ln L_c$,[2-5] reflecting a gradual ballistic-to-diffusive crossover rather than a sharp transition.

Most previous studies, however, have identified these ln-scale trends only at the level of ensemble-averaged quantities, such as $K(k)$ or $K(L_c)$ obtained after summation over all phonon modes. In such analyses, the logarithmic dependence is interpreted as a collective consequence of the full MFP spectrum: as the length scale increases, additional groups of long-MFP phonons gradually become involved in heat conduction, and the accumulation of their contributions produces an effective $K$ that grows as $\ln L_c$ or varies as $\ln(1/k)$. This figure illustrates how MFP phonons contribute to the slow convergence of the thermal conductivity $K$ with length. However, because it integrates over the detailed response of individual phonon modes, it does not reveal whether any single mode by itself obeys a logarithmic suppression law, especially in realistic device geometries.

Previous work by Vermeersch and Mingo[6] systematically analyzed quasiballistic heat removal from radially symmetric nanoscale heat sources in crystalline solids using a multidimensional Boltzmann transport framework based on first-principles phonon properties. For crystalline silicon, they showed that the apparent thermal conductivity $K_{\mathrm{eff}}$ experienced by a hotspot of characteristic radius $R$ exhibits a strong size dependence and, in the intermediate quasiballistic regime between the ballistic ($K_{\mathrm{eff}} \propto R$) and diffusive ($K_{\mathrm{eff}} \approx K_{\mathrm{bulk}}$) limits, follows an approximately logarithmic scaling $K_{\mathrm{eff}} \propto \ln R$. However, their analysis is essentially mode-averaged: the detailed mode-dependent phonon properties obtained from first principles are condensed into an effective propagator and a scalar suppression function $S(R)$, so that the contributions of individual phonon branches to the logarithmic $K_{\mathrm{eff}} \sim \ln R$ behavior in crystalline silicon are not resolved explicitly.

In this work, we address this gap by resolving phonon transport into mode-by-mode contributions and examining how each contributes to quasiballistic scaling in silicon. We formulate a modal-weighting framework in the spectral MFP domain that embeds known bulk ln-scaling into a Si-layer-specific spectral picture. A semianalytic treatment (supplementary material Sec. S1) shows that film confinement fundamentally alters the modal behavior: only quasiballistic modes with MFPs comparable to the Si layer thickness retain logarithmic accumulation, whereas fully diffusive modes and fully ballistic long-MFP modes yield saturated, nonlogarithmic contributions. In other words, the ln-regime is predicted to be spectrally selective rather than universal across the entire phonon spectrum.

Because a direct experimental demonstration of this modal ln-scaling in thin Si layers is currently impractical, we instead seek validation in a realistic silicon-on-insulator (SOI) transistor geometry whose thermal behavior has been quantitatively benchmarked against both experiments and prior full-dispersion BTE simulations. A full phonon dispersion model (FPDM) is used to solve the spectral Boltzmann transport equation (BTE) for a Si device layer subjected to a localized $10 \times 10\ \mathrm{nm}^2$ heat source. To quantify modal contributions under such nonlocal conditions, we introduce the weighted thermal transport resistance (WTRT)—the product of modal thermal conductivity and modal thermal resistance—which captures the realized heat-carrying role of each phonon mode more faithfully than modal conductivity alone. By analyzing the WTRT for low-frequency (LF) and high-frequency (HF) phonon groups as the transport regime evolves from ballistic to quasiballistic, we demonstrate that the ln-scale behavior predicted by the semianalytic framework indeed emerges in the confined Si layer of the SOI transistor. Our results thereby connect fundamental ln-type quasiballistic physics to device-level heat transport in advanced silicon technologies.

## II. SEMIANALYTIC FRAMEWORK FOR MODE-RESOLVED QUASIBALLISTIC TRANSPORT

To investigate mode-resolved logarithmic quasiballistic heat transport in silicon, we employ a semianalytic framework, described in detail in the supplementary material (Sec. S1.1–S1.4). This framework is formulated directly in the spectral MFP domain: instead of introducing a single "effective" phonon mode or a mode-averaged propagator, we explicitly retain the full dependence of the thermal conductivity spectrum $K(\Lambda)$, the modal transport weight $w(\Lambda)$, and the nonlocal suppression kernel $S(\Lambda/\sigma)$ on the phonon MFP ($\Lambda$) and on the characteristic nonlocal length $\sigma$ (e.g., film thickness or hotspot radius). Figure 1 serves as a visual roadmap for this Section-II discussion: Fig. 1(a) sketches the approximately log-uniform conductivity plateau of crystalline Si, Fig. 1(b) summarizes the corresponding continuum-scale regime map of the plateau-aggregated response as $\sigma$ varies, and Fig. 1(c) provides a compact flowchart linking $K(\Lambda)$, $w(\Lambda)$, and $S(\Lambda/\sigma)$ to the $K$-weighted accumulation and its ln-active versus saturating contributions. Here, $\Lambda$ denotes the phonon MFP at angular frequency $\omega$ ($\Lambda \equiv \Lambda_{\omega}$), and we hereafter drop the subscript and simply write $\Lambda$ in the semianalytical derivations that follow. Within this spectral picture, all logarithmic signatures arise from the accumulation of properly weighted modal contributions over the MFP distribution rather than from any phenomenological fitting of an effective nonlocal conductivity.

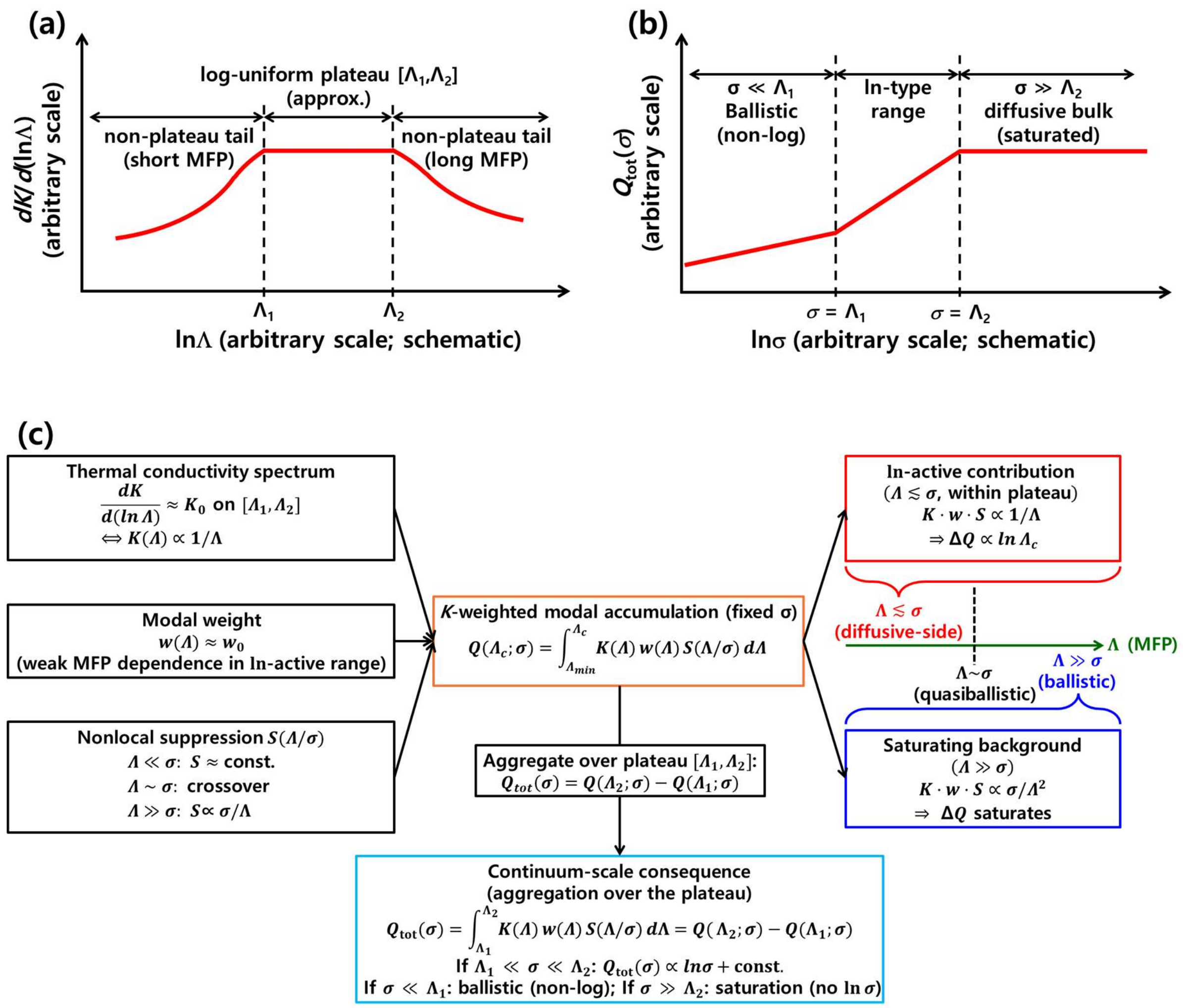

**FIG. 1.** Schematic roadmap for Sec. II (definitions and derivations in Supplementary Material Secs. S1.1–S1.4). (a) Log-uniform conductivity plateau of crystalline Si over $[\Lambda_1, \Lambda_2]$ in MFP space (schematic; non-plateau tails not to scale). (b) Plateau-aggregated response versus $\sigma$: a ln-type dependence appears only for $\Lambda_1 \ll \sigma \ll \Lambda_2$. (c) Flowchart linking the spectrum, modal weight, and suppression kernel to the fixed-$\sigma$ accumulation $Q(\Lambda_c; \sigma)$ and the $\sigma$-dependent plateau-aggregated response $Q_{\text{tot}}(\sigma)$, separating ln-active ($\Lambda \lesssim \sigma$) and saturating background ($\Lambda \gg \sigma$) contributions.

The starting point is the well-known result that crystalline Si exhibits an approximately log-uniform conductivity spectrum over a broad MFP interval $[\Lambda_1, \Lambda_2]$, such that each decade of $\Lambda$ contributes comparably to the bulk thermal conductivity.[7-11] Quantitatively, this logarithmic plateau can be expressed as an approximately constant conductivity per logarithmic MFP interval,

$$\frac{dK}{d(\ln\ \Lambda)} \approx K_0\ (\Lambda_1 \lesssim \Lambda \lesssim \Lambda_2), \qquad (1)$$

which is equivalently written, in terms of the spectrum per unit MFP, as

$$K(\Lambda) \equiv \frac{dK}{d\Lambda} \approx \frac{K_0}{\Lambda}, (\Lambda_1 \lesssim \Lambda \lesssim \Lambda_2). \qquad (2)$$

Here, $K_0$ denotes the approximately constant value of the thermal conductivity spectrum per logarithmic MFP interval, i.e., the plateau of $dK/d(\ln\ \Lambda)$ in Eq. (1).[7-11] We refer to this logarithmic plateau as the range of $\Lambda$ over which the spectrum is approximately constant when expressed as a function of $\ln \Lambda$, with $\Lambda_1$ and $\Lambda_2$ denoting the lower and upper bounds of the plateau, respectively. Near a nanoscale hotspot or confined heat source, this spectrum is combined with a mode-dependent driving factor, represented by $w(\Lambda)$, and with a geometry-dependent suppression kernel $S(\Lambda/\sigma)$ that encodes how confinement and nonlocality reduce the contribution of each mode as its MFP becomes comparable to or larger than $\sigma$.

The resulting $K$-weighted modal quantity $Q(\Lambda_c)$ is defined by integrating $K(\Lambda)$ $w(\Lambda)$ $S(\Lambda/\sigma)$ up to a cutoff MFP $\Lambda_c$. In the bulk-like diffusive-side regime ($\Lambda \ll \sigma$), where $w(\Lambda)$ and $S(\Lambda/\sigma)$ vary only weakly with $\Lambda$, the integrand acquires an effective $1/\Lambda$ structure, and $Q(\Lambda_c)$ grows as $\ln \Lambda_c$. This establishes that, at fixed $\sigma$, diffusive-side modes generate a logarithmic accumulation with respect to the MFP but do not yet specify how any effective nonlocal conductivity depends on the geometric length $\sigma$ itself; that $\sigma$-dependence is instead obtained from the plateau-aggregated quantity $Q_{\text{tot}}(\sigma)$ [see Fig. 1(b) and Supplementary Material Sec. S1.4].

The same modal framework is then extended to thin Si layers. For quasiballistic modes with $\Lambda$ comparable to the film thickness $L$ (i.e., $\Lambda \sim L$), the kernel remains of order unity and only weakly $\Lambda$ dependent, so the $K$-weighted modal accumulation in this regime also scales as $\ln \Lambda/L$, continuously extending the bulk diffusive-side logarithmic trend up to the crossover scale set by $L$. In contrast, fully ballistic long-MFP modes with $\Lambda \gg L$ experience a suppression kernel that decays asymptotically as $S(\Lambda/L) \propto L/\Lambda$. This modifies the integrand from $1/\Lambda$ to $1/\Lambda^2$ and causes the ballistic contribution to saturate to a finite constant with no $\ln \Lambda$ term. Thus, the semianalytic analysis shows that logarithmic accumulation is produced only by modes with $\Lambda \lesssim \sigma$ (diffusive-side and quasiballistic), whereas modes with $\Lambda \gg \sigma$ contribute a nonlogarithmic, saturated background.

To connect these modal results to the continuum-scale nonlocal conductivity $K_{\text{eff}}(\sigma)$, the supplementary material (Sec. S1.4) introduces an aggregated spectral quantity $Q_{\text{tot}}(\sigma)$, defined by integrating $K(\Lambda)$ $S(\Lambda/\sigma)$ over the logarithmic plateau region $[\Lambda_1,\Lambda_2]$. When the nonlocal length lies inside this plateau ($\Lambda_1 \ll \sigma \ll \Lambda_2$), the integral can be decomposed into a "diffusive-side + quasiballistic" part ($\Lambda < \sigma$) and a fully ballistic part ($\Lambda > \sigma$). The former retains the $1/\Lambda$ kernel and yields $Q_{tot}(\sigma) \propto \ln \sigma$, whereas the latter saturates to a constant such that $K_{eff}(\sigma)$ inherits a $\ln \sigma$ dependence only in this intermediate quasiballistic range. Outside the plateau, the logarithmic behavior disappears: in the fully ballistic limit ($\sigma \ll \Lambda_1$) and in the fully diffusive bulk limit ($\sigma \gg \Lambda_2$), all contributing modes lie on the same side of the $\Lambda \sim \sigma$ boundary, the suppression kernel effectively becomes constant over the active spectrum, and $Q_{tot}(\sigma)$ [hence $K_{eff}(\sigma)$] becomes independent of $\sigma$.

In summary, the semianalytic framework provides a transparent, spectral interpretation of the ln-type quasiballistic regime in terms of mode-resolved quantities. This shows that logarithmic size dependence in Si arises from a specific subset of modes—those whose MFPs lie within a log-uniform conductivity plateau and are diffusive-side or quasiballistic with respect to the relevant length scale $\sigma$—while fully ballistic long-MFP modes contribute only a saturated background. Because a direct experimental test of these modal predictions in thin Si device layers is not currently feasible, the numerical results in Sections IV–V specialize in this framework to the weighted thermal transport resistance used in this work and validate the predicted modal ln-scaling via full-dispersion BTE simulations (FPDM) of a nanoscale hotspot in a thin-film SOI transistor.

## III. PROBLEM SETUP AND DEFINITIONS: MODAL NONLOCALITY PARAMETER

Nonlocal phonon transport is governed by the competition between the characteristic propagation length of a phonon mode and the spatial scale of the driving temperature gradient. To quantify this effect at the mode-resolved level, we introduce the modal nonlocality parameter $\xi_\lambda = -\nabla T_\lambda / \Lambda_{f,\lambda}$, where $T_\lambda$ is the modal temperature[12-18] and where $\Lambda_{f,\lambda}$ denotes the phonon MFP of mode $\lambda$ in the thin film. Here, the phonon MFP in the thin film, $\Lambda_{f,\lambda}$, is related to the bulk MFP $\Lambda_{b,\lambda}$ through a mode-dependent suppression function $\phi_{\lambda}$, such that $\Lambda_{f,\lambda} = \phi_{\lambda}\Lambda_{b,\lambda}$. The form of $\phi_{\lambda}$ and its implementation for in-plane thermal transport in silicon thin films follow Ref.[19], and the detailed expression and band-resolved values are summarized in Sec. S2 of the supplementary material (see Fig. S1).

Boundary scattering was assumed to be fully diffuse,[20] corresponding to a specularity parameter of zero, where $\nabla T_{\lambda}$ is the modal temperature gradient and where $(\nabla T_{\lambda})^{-1}$ denotes the local modal temperature-variation length. The parameter $\xi_{\lambda}$ increases monotonically with the degree of ballisticity experienced by mode $\lambda$ and therefore serves as a direct measure of modal suppression in the quasiballistic regime.

This definition is physically consistent with the traditional nonlocality arguments embedded in spectral kernels,[11,21] which depend on the ratio between the phonon MFP and the characteristic scale of the temperature field. The key distinction lies in the fact that $\xi_{\lambda}$ is defined for each phonon mode, allowing explicit evaluation of mode-dependent nonequilibrium near nanoscale hotspots, where different phonon branches experience different gradients and scattering pathways. Thus, $\xi_{\lambda}$ provides an interpretable, mode-resolved representation of the same underlying nonlocal physics encoded in MFP-based kernels but with specific relevance to modal temperature gradients and suppression behavior.

To quantify the response of individual phonon modes to nonlocal thermal driving, including mode-by-mode phonon scattering processes, we define the modal thermal resistance $R_{\lambda}$ and modal thermal conductivity $K_{\lambda}$ as

$$R_{\lambda} = \frac{\Delta T_{\lambda}}{q_{\lambda}}, \qquad K_{\lambda} = -\frac{q_{\lambda}}{\nabla T_{\lambda}}, \qquad (3)$$

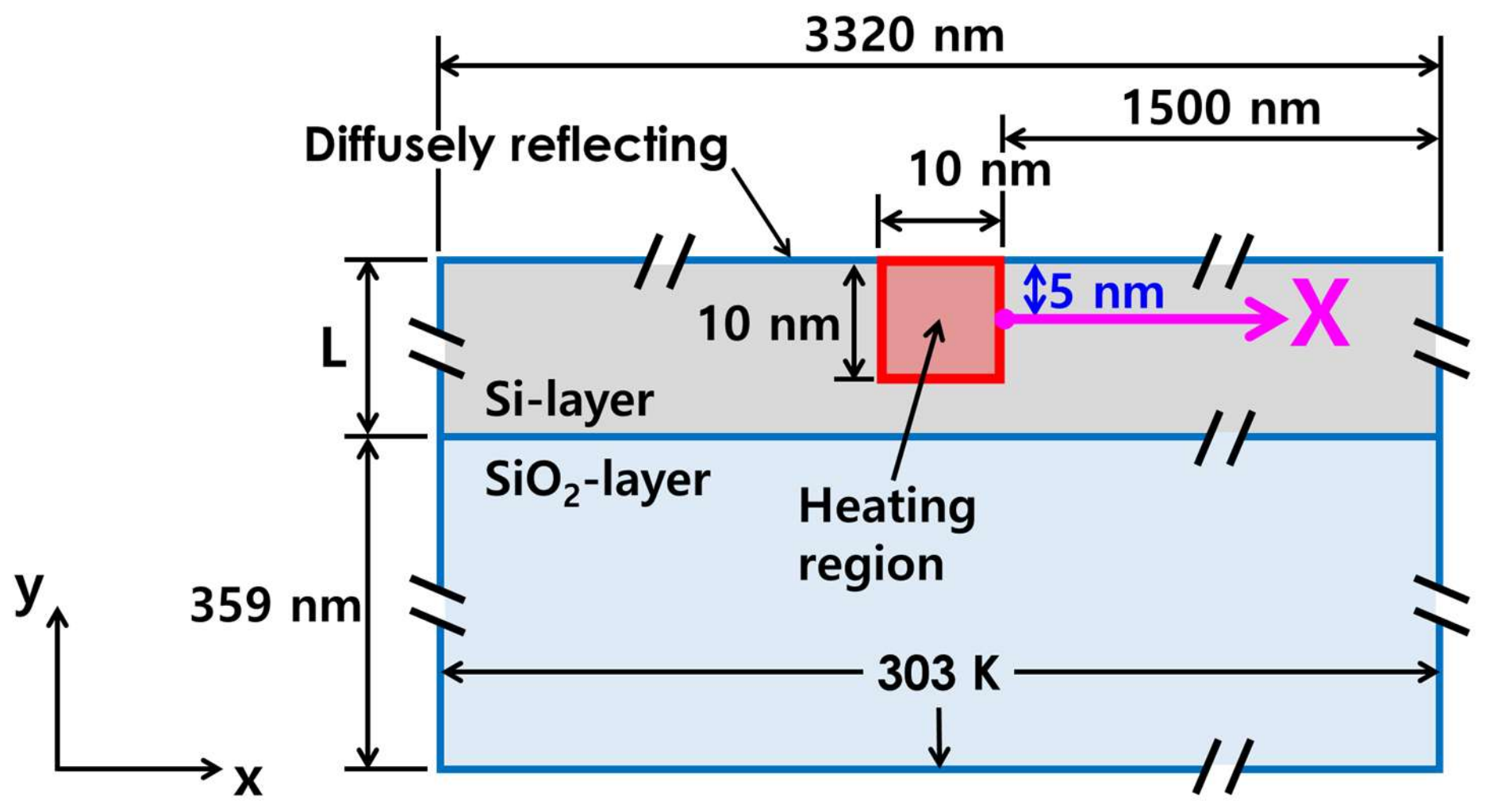

**FIG. 2**. The computational domain used in this study is identical to that presented in Ref.[33], which was constructed on the basis of the experimental geometry of Goodson *et al*.[31]. The analysis in this work is carried out along the in-plane *X*-direction indicated by the magenta arrow, starting from the edge of the heated region.

where $q_\lambda$ is the heat flux carried by mode $\lambda$, defined below in Eq. (6), Here, $\Delta T_\lambda$ and $\nabla T_\lambda$are evaluated from the mode-resolved temperature field $T_\lambda(\mathbf{r})$, which is generally nonequilibrium near the nanoscale hotspot (i.e., $T_\lambda$ can differ from the lattice temperature $T_L$ and from mode to mode). In other words, Eq. (3) does not use a single equilibrium/lattice temperature field for all modes; rather, the nonequilibrium enters through the mode dependence of both $T_\lambda$ (hence $\Delta T_\lambda$and $\nabla T_\lambda$) and $q_\lambda$ obtained from the BTE solution. In this formulation, $R_\lambda$ is defined as the temperature drop associated with mode $\lambda$ per unit heat flux it transports, whereas $K_\lambda$ characterizes its effective heat-carrying response under the same (mode-resolved) driving field. The product, $K_\lambda R_\lambda$, serves as a modal participation factor that quantifies the realized contribution of mode $\lambda$ to macroscopic heat conduction after accounting for nonlocal suppression. Note that $K_\lambda R_\lambda = -\frac{\Delta T_\lambda}{\nabla T_\lambda}$, so $K_\lambda R_\lambda$ is governed directly by the spatial profile of the nonequilibrium modal temperature $T_\lambda$. In the local-equilibrium (diffusive) limit, $T_\lambda \to T_L$ for all modes and Eq. (3) reduces to the usual Fourier-form definitions.

## IV. NUMERICAL VALIDATION IN A REALISTIC SOI TRANSISTOR GEOMETRY

We next validate the semianalytical predictions using a full-phonon-dispersion Boltzmann transport model (FPDM)[22,23] applied to a realistic SOI transistor structure (Fig. 2), which resolves mode-resolved scattering processes among individual phonon modes. Direct, mode-resolved experimental measurements of quasiballistic ln-scaling in thin Si layers are, at present, essentially infeasible because they would require resolving modal heat fluxes and temperature drops around a deeply scaled hotspot. Instead, we rely on a spectral BTE solver that has been quantitatively benchmarked against both gate-thermometry measurements and prior full-dispersion BTE calculations for SOI devices. By applying this FPDM to the Si device layer containing a $10 \times 10$ $\mathrm{nm}^2$ hotspot, we obtain mode-resolved temperatures, temperature gradients, and transport coefficients, which are then used to evaluate the quantities appearing in the semianalytic framework and to test the predicted modal ln-scaling.

### A. Spectral Boltzmann transport equation (BTE)

For each phonon mode $\lambda \equiv$ (branch, $\omega$) and propagation direction $\mathbf{s}$, the deviational energy density $e_\lambda(\mathbf{r}, \mathbf{s})$ satisfies the following equation derived from the BTE:[22,23]

$$\frac{\partial e_\lambda}{\partial t} + v_\lambda \cdot \nabla e_\lambda = \frac{e_\lambda^0(T_\lambda) - e_\lambda}{\tau_\lambda} + \dot{H}, \qquad (4)$$

where **r** is the position vector, branch denotes the phonon polarization, $\omega$ is the angular frequency, $v_\lambda$ is the group velocity (for optical phonons, $v_\lambda$ is assumed to be zero owing to their negligible group velocity in silicon),[22,23] $\tau_\lambda(\omega)$ is the mode-dependent relaxation time, and $e_\lambda^0(T_\lambda)$ is the equilibrium energy density. The full dispersion and scattering model, which includes longitudinal acoustic (LA), transverse acoustic (TA), and optical branches with a 6–6–1 band discretization, follows the work of Narumanchi *et al.*[22,23] Here, $\dot{H}$ is an explicit volumetric source term (W/m$^3$) that accounts for the energy transferred from electrons to phonons via electron–phonon scattering within the hotspot region. The total volumetric heat generation rate is denoted by $q_{\mathrm{vol}}$ (set to $10^{18}$ W/m$^3$ in this study). To represent heating partition between acoustic and optical sectors, we introduce the acoustic heating fraction $qa \in [0,1]$ and define the sector source strengths as $\dot{H}_A = qa\, q_{\mathrm{vol}}$, $\dot{H}_O = (1 - qa)\, q_{\mathrm{vol}}$, where $\dot{H}_A$ applies when Eq. (4) is solved for the acoustic modes (LA and TA), and $\dot{H}_O$ applies when Eq. (4) is solved for the optical modes.

## B. FPDM: Modal temperature, temperature gradient, and temperature difference

We adopt the FPDM formulation of Narumanchi *et al.*,[22,23] in which the longitudinal and transverse acoustic branches are partitioned into discrete bands (LA1–LA6 and TA1–TA6). Each band is treated as an independent energy carrier with a bandwise specific heat, density of states, group velocity, and band-averaged relaxation time. The complete set of bandwise parameters is provided in Table I. The effective relaxation times ($\tau_{\mathrm{eff},\lambda}$) listed in Table I are obtained by combining the mode-by-mode relaxation times reported in Ref.[23] via Matthiessen's rule. The phonon density of states (DOS) for the transverse acoustic (TA) and longitudinal acoustic (LA) branches is constructed by fitting the DOS reported in Ref.[24] so that its absolute magnitude matches that reported in Ref.[25] and then performing bandwise averaging. The values of $C_\lambda$ and $v_\lambda$ are taken from Ref.[26].

**TABLE I**. Bandwise phonon properties used as input data in this study.

| Phonon mode | Frequency range (THz) | $\omega_\lambda$ ($10^{12}$ rad/s) | $C_\lambda$ (J/kg·K) | $v_\lambda$ (m/s) | $\tau_{\mathrm{eff},\lambda}$ (ps) | $\Lambda_{b,\lambda}$ (nm) | DOS ($10^{15}$ s/rad·m$^3$) |
|---|---|---|---|---|---|---|---|
| LA1 | 0–1.98 | 6.2167 | 582.0 | 9147.8 | 1305.2 | 11939.7 | 0.3136 |
| LA2 | 1.98–3.96 | 18.65 | 4497.6 | 8559.9 | 41.5 | 355.4 | 0.3354 |
| LA3 | 3.96–5.94 | 31.0833 | 15054.3 | 7459.1 | 24.8 | 184.6 | 0.3596 |
| LA4 | 5.94–7.92 | 43.5167 | 35587.3 | 6503.8 | 19.1 | 124.2 | 0.4365 |
| LA5 | 7.92–9.89 | 55.95 | 79411.9 | 5398.2 | 16.9 | 91.2 | 1.4504 |
| LA6 | 9.89–11.87 | 68.3833 | 184226.8 | 3907.5 | 8.2 | 31.9 | 8.0506 |
| TA1 | 0–0.74 | 2.3213 | 314.5 | 5329.6 | 6215.6 | 33126.7 | 0.0719 |
| TA2 | 0.74–1.48 | 6.9640 | 2071.1 | 5523.3 | 969.3 | 5353.8 | 0.0969 |
| TA3 | 1.48–2.22 | 11.6067 | 5424.7 | 4976.7 | 398.6 | 1983.9 | 0.1982 |
| TA4 | 2.22–2.96 | 16.2493 | 18289.9 | 4122.1 | 59.5 | 245.1 | 3.4334 |
| TA5 | 2.96–3.69 | 20.8920 | 55692.0 | 2926.7 | 32.8 | 96.0 | 2.5447 |
| TA6 | 3.69–4.43 | 25.5347 | 580061.4 | 1315.2 | 7.2 | 9.5 | 0.2636 |

The modal temperature $T_\lambda$ is computed directly from the full phonon distribution moment:

$$e_\lambda = \int_{T_{\mathrm{ref}}}^{T_\lambda} C_\lambda(T)\, dT\,, \qquad (5)$$

where $C_\lambda$ is the phonon modal specific heat, here approximated as a constant for temperatures above 300 K.[22,23] The lattice temperature, $T_L$, can now be calculated via the total phonon energy ($e_{\mathrm{total}}$), defined as $e_{total} = \sum_{i=1}^{N_{bands}} e_{\lambda} = \int_{T_{ref}}^{T_L} C_{tot}\, dT$, where $N_{\text{bands}}$ is the total band number (i.e., LA mode 6 bands, TA mode 6 bands, and optical mode 1 band) and $C_{tot}$ is the total specific heat of the silicon. We choose $T_{\text{ref}}$ = 303 K to coincide with the isothermal boundary temperature used in the FPDM–BTE simulations (Fig. 2). According to previous studies, 20.4% of the thermal energy ($q_{\text{vol}}$) generated by excited electrons is transferred to the acoustic mode (AM), whereas the remaining 79.6% is transferred to optical mode (OM).[27,28] Furthermore, the energy delivered to AM is distributed nearly uniformly across different phonon modes in the Si layer.[29] In this study, we therefore assume that 20% of the thermal energy generated by excited electrons is transferred to AM, whereas the remaining 80% is delivered to OM, with the AM energy equally distributed among six phonon bands.

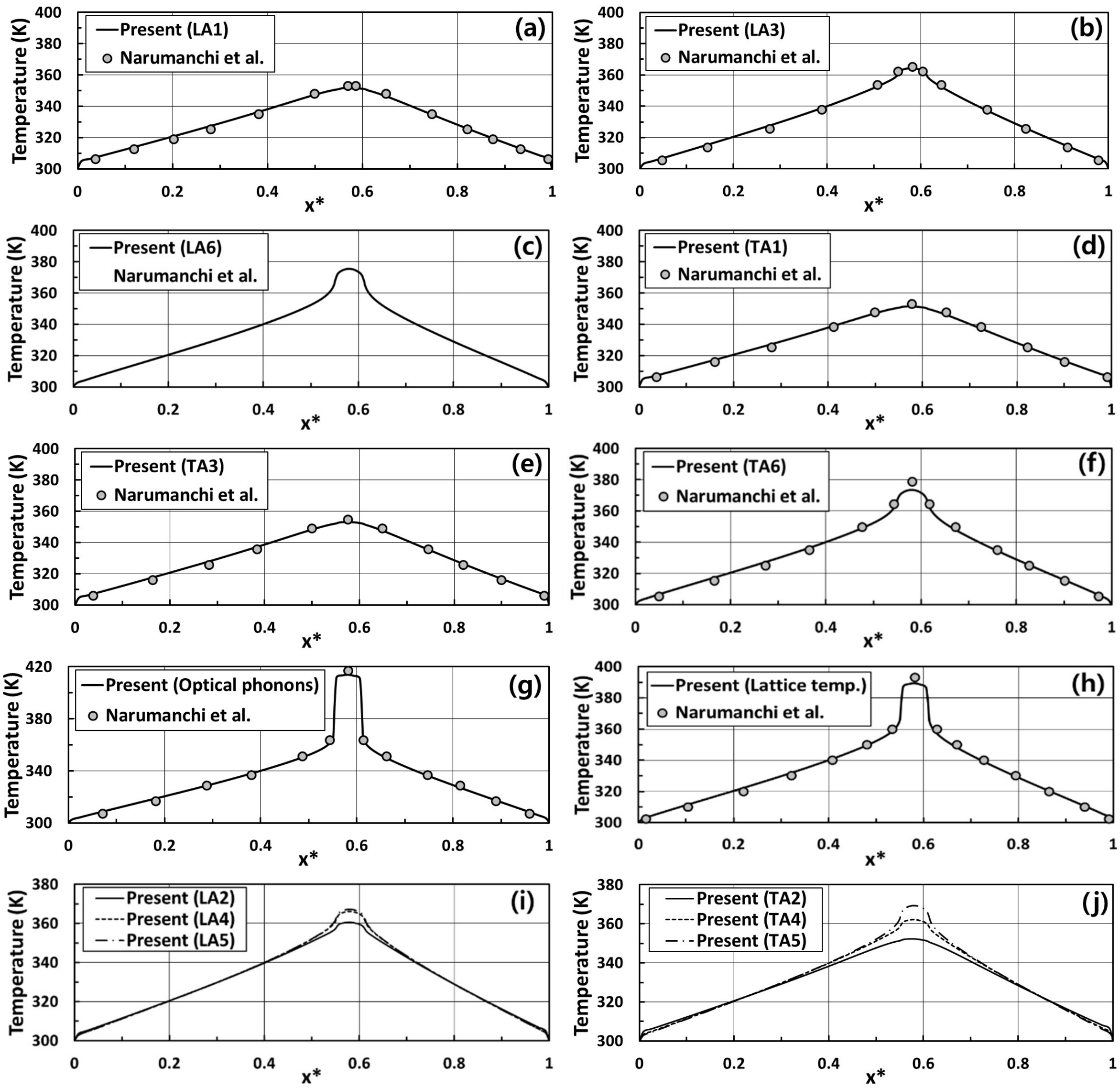

**FIG. 3**. Benchmark comparison between the calculated temperature distributions obtained with the present FPDM solver and those reported by Narumanchi *et al.*[23] for $qa = 0$, using the SOI transistor case and thermal boundary conditions described in Ref.[23], showing good agreement within the error margins: (a) LA1, (b) LA3, (c) LA6, (d) TA1, (e) TA3, (f) TA6, (g) optical phonons, and (h) Lattice temperature ($T_L$). Panels (i) LA2, LA4, and LA5 and (j) TA2, TA4, and TA5 show only our results, because Narumanchi *et al.*[23] did not report these cases. The *x*-axis denotes the dimensionless coordinate $x^* = X/1500$, where $X$ is measured along the in-plane channel ($X$) direction starting from the edge of the heated region (see Fig. 2 for the coordinate definition).

The accuracy of the simulated temperature distributions is critical for estimating $\Delta T_{\lambda}$ and $\nabla T_{\lambda}$ in Eq. (3). To verify the quantitative accuracy of our in-house FPDM implementation, we benchmarked our solver against the published FPDM–BTE results of Narumanchi et al.[23] by reproducing their reported SOI transistor case at $qa = 0$, using the same device configuration, heat-generation condition, and thermal boundary conditions. As shown in Fig. 3, our computed mode-resolved and lattice-temperature profiles agree well with those reported in Ref.[23], with a maximum (relative) deviation of 1.41%. Having established this level of agreement for the Ref.[23] configuration, we subsequently apply the validated FPDM solver to the SOI transistor geometry considered in this work (Fig. 2), which differs from the benchmark configuration in Ref.[23].

Considering the heat transfer process from nanoscale hotspots, high-energy phonons such as optical phonons are emitted from hotspots. Because of energy and momentum conservation, the phase space for these phonons to scatter with low-frequency phonons is extremely limited, and they instead predominantly scatter with phonon modes of relatively high energy.[23] Motivated by this, in the present study, we classify the phonon modes into low-frequency (LF) and high-frequency (HF) groups. The LF group consists of the LA1 and TA1–TA3 phonon modes, which have no direct scattering channel with the optical phonon modes emitted from the NH and therefore do not receive heat directly from the hotspot. The HF group consists of the LA2–LA6 and TA4–TA6 phonon modes, which can scatter directly with the optical phonon modes and thus receive heat directly from the hotspot.

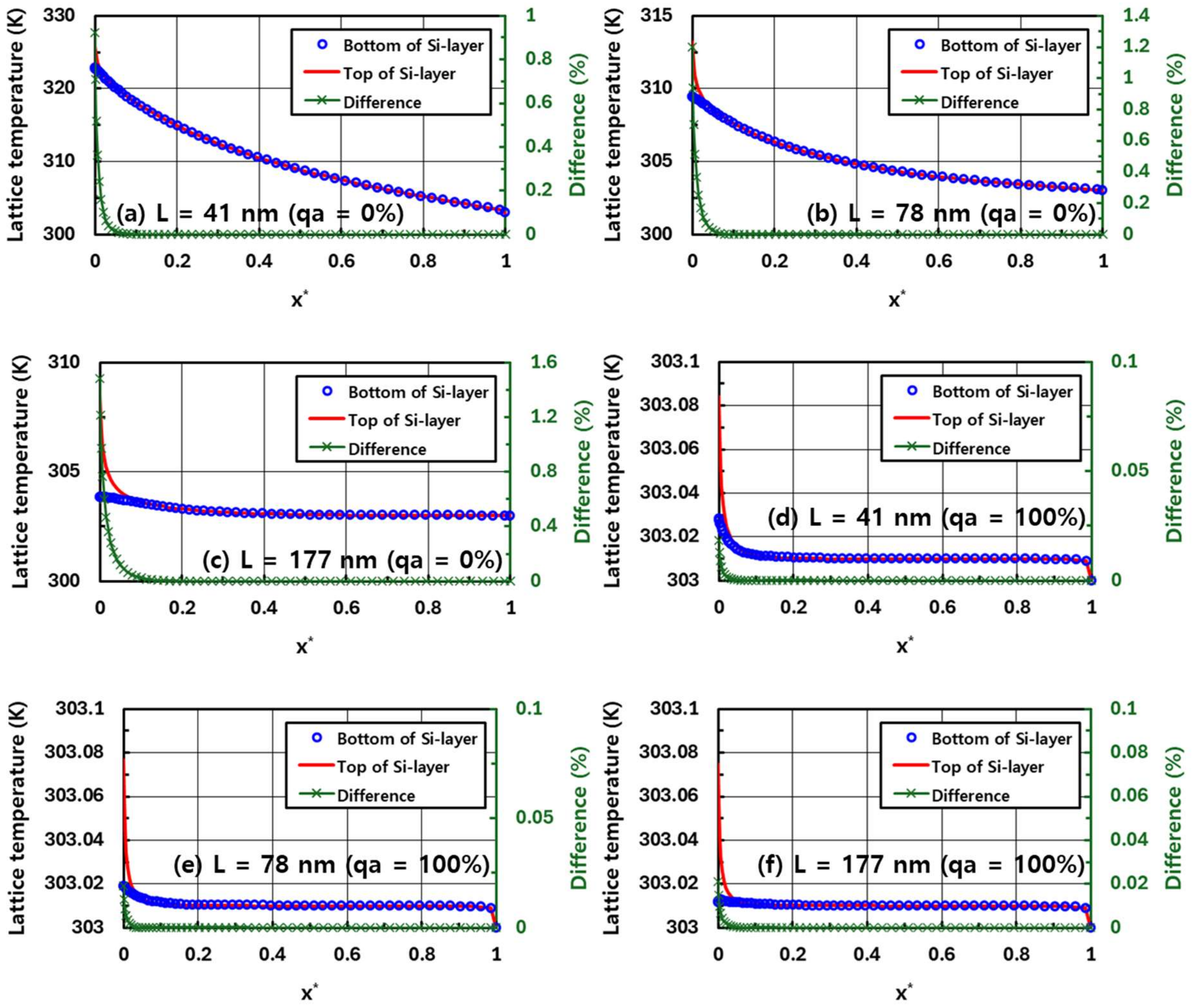

**FIG. 4**. Lattice temperature distributions as a function of $x^*$ (= $X$/1500, see Fig. 2) in the Si layer for $qa$ = 0% and 100%, validating the one-dimensional phonon transport assumption: (a) $L$ = 41 nm ($qa$ = 0%), (b) $L$ = 78 nm ($qa$ = 0%), (c) $L$ = 177 nm ($qa$ = 0%), (d) $L$ = 41 nm ($qa$ = 100%), (e) $L$ = 78 nm ($qa$ = 100%), (f) $L$ = 177 nm ($qa$ = 100%).

Because the buried $SiO_2$ layer has a much lower thermal conductivity than Si, phonon transport in the Si device layer of the SOI transistor is expected to occur predominantly along the channel ($x$) direction.[23] Before adopting a one-dimensional (1D) description of phonon transport along $x$ in our analysis, we therefore examine whether out-of-plane ($y$-direction) heat spreading within the thin Si layer is negligible for the device geometries of interest. To this end, Fig. 4 compares the lattice temperature ($T_L$) distributions at the top and bottom surfaces of the Si layer for $qa$ = 0% and $qa$ = 100% and for layer thicknesses $L$ = 41 nm, 78 nm, and 177 nm. For all three thicknesses, $T_L$ is lower when $qa$ = 100% than when $qa$ = 0%, which is consistent with the fact that supplying $q_{\mathrm{vol}}$ to the acoustic mode (AM) leads to more efficient energy transport owing to the high group velocity of the AM. As $L$ increases, the overall $T_L$ decreases and the top-surface temperature profile, on which the hotspot is imposed, exhibits a sharper decay along $x$, whereas the bottom-surface profile varies more gradually because it is farther from the hotspot. A noticeable enhancement of $y$-directional transport is observed only in the immediate vicinity of the hotspot, where the thin Si layer allows some cross-plane heat flow. However, even in the most unfavorable case ($L$ = 177 nm, $qa$ = 0%), the maximum difference between the top- and bottom-surface temperatures remains below 1.6%. These results demonstrate that, although local two-dimensional effects exist near the hotspot, phonon transport in the Si layer can be treated, to a very good approximation, as one-dimensional along $x$ in the subsequent analysis. Under this 1D approximation, the heat flux in the $x$-direction is given by:[30]

$$q_\lambda = \int_{\omega_\lambda} v_\lambda f_\lambda^{eq}(\omega)\, D(\omega)\, \hbar\omega\, d\omega, \qquad (6)$$

where $D(\omega)$ is the density of states, $f_\lambda^{eq}$ is the equilibrium Bose–Einstein distribution function for mode $\lambda$, and $\hbar$ is the reduced Planck's constant.

## C. Geometry, boundary conditions, and numerical scheme

As shown in Fig. 2, thin-film silicon layers with thicknesses $L$ = 41, 78, and 177 nm are simulated via two-dimensional finite-volume discretization combined with angular quadratures over the unit sphere. The top boundary is modeled as diffusely reflecting and adiabatic,[23] whereas all remaining boundaries are held at an isothermal temperature $T_0$ = 303 K. Steady-state solutions are obtained by iterating Eq. (4) until all the modal energy densities converge.

By fixing the hotspot size (10 × 10 nm$^2$), the source geometry and phonon phase space remain constant, allowing a direct evaluation of how reduced modal specific heat and limited scattering channels enhance local temperature gradients and effective thermal resistance. The selected dimension also reflects realistic nanoscale power dissipation regions in modern semiconductor devices, where the hotspot size is comparable to that of typical phonon MFPs. The computational grid and angular resolution were set to ensure numerical convergence (< 0.1% variation) and to capture the steep spectral gradients near the hotspot. As mentioned earlier, Table I lists the angular frequencies of these modes and provides all the input parameters required to solve the phonon BTE and compute the modal temperature and modal heat flux in Eq. (3).

## D. Validation of the FPDM solver

The FPDM used in this study is an in-house spectral BTE solver developed by the authors' group. Its quantitative accuracy has been established through a series of benchmarks against both experiments and independent full-dispersion BTE calculations. First, Goodson *et al*.[31] reported gate thermometry measurements of SOI transistor channel temperatures; our FPDM reproduces these measurements with excellent agreement over the relevant power and geometry ranges. Second, the solver reproduces the full-distribution BTE results previously obtained by one of the present authors (Jin) and Lee[32,33] for local heating and electrostatic-discharge (ESD) scenarios in SOI structures, with deviations typically within 1~2% in both the steady and transient regimes.

While first-principles studies commonly report mode-resolved intrinsic lifetimes, the FPDM of Narumanchi et al. additionally requires a band-to-band energy-exchange (interaction) relaxation-time matrix, $\tau_{ij}$, to describe interband inelastic coupling under the adopted band discretization (see Table 2 of Ref.[23]). Because $\tau_{ij}$ cannot be inferred from published lifetime spectra alone, a fully first-principles re-parameterization of the FPDM would require additional scattering-operator post-processing and coarse-graining onto the FPDM band basis, which is beyond the scope of the present work. We therefore retain the validated baseline FPDM inputs and present Fig. 5 as a current benchmark against the literature: a bulk-only comparison of the acoustic-phonon relaxation-time inputs used in this work ($T$ = 300 K) with representative first-principles/ALD (anharmonic lattice dynamics) datasets (Esfarjani et al.[34]; Sellan et al.[35]). As shown in Fig. 5, the relaxation-time spectra adopted here fall within the envelope spanned by these literature results at temperatures near room temperature, supporting the suitability of the baseline FPDM inputs for the present model validation.

Taken together with the device-level validations discussed above, these benchmarks indicate that the FPDM captures the relevant phonon-transport physics in SOI devices and that the trends reported in the present simulations are governed by intrinsic modal behavior rather than numerical artifacts. Consequently, the FPDM-based results provide a reliable surrogate for mode-resolved experimental data that are not currently accessible and form a solid basis for testing the semianalytic predictions of ln-scale quasiballistic transport in thin Si layers.

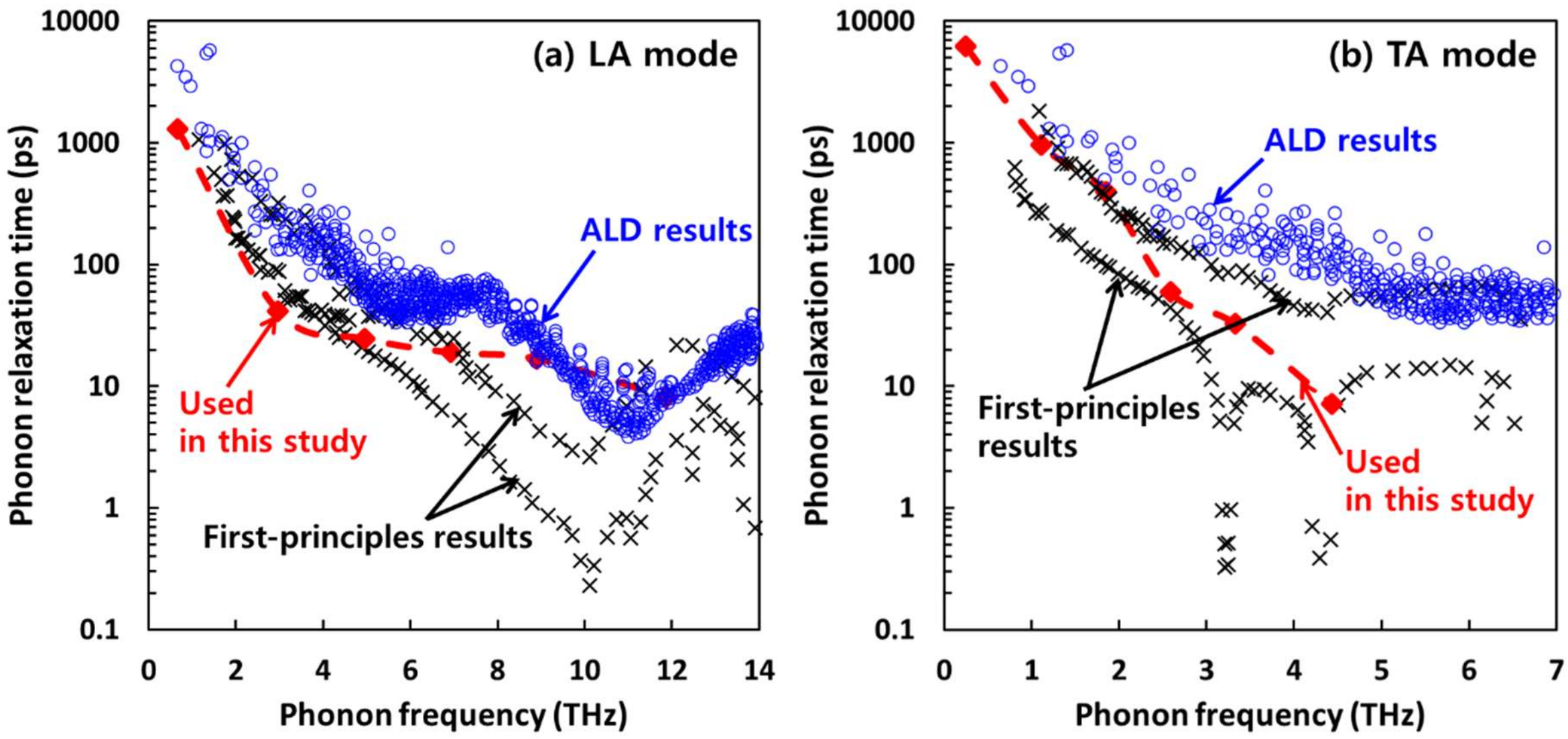

**FIG. 5**. Bulk acoustic-phonon relaxation times used in the present FPDM input set ($\tau_{\mathrm{eff},\lambda}$ in Table I), compared with representative first-principles results (Esfarjani et al.[34]) and ALD (anharmonic lattice dynamics) results (Sellan et al.[35]) near room temperature. For Esfarjani et al.[34], the total lifetime is constructed from the reported normal ($\tau_N$) and Umklapp ($\tau_U$) components using Matthiessen's rule, $1/\tau = 1/\tau_U + 1/\tau_N$. Sellan et al.[35] report combined LA+TA phonon–phonon lifetimes; these data are shown here as a literature reference for both (a) LA and (b) TA modes. Because the Esfarjani et al.[34] dataset exhibits noticeable scatter, we plot its upper and lower envelopes. The datasets of Esfarjani et al.[34] and Sellan et al.[35] were obtained via the extraction procedure of Ref.[36].

## V. RESULTS AND DISCUSSION

### A. Modal ln-scaling behavior across LF and HF modes

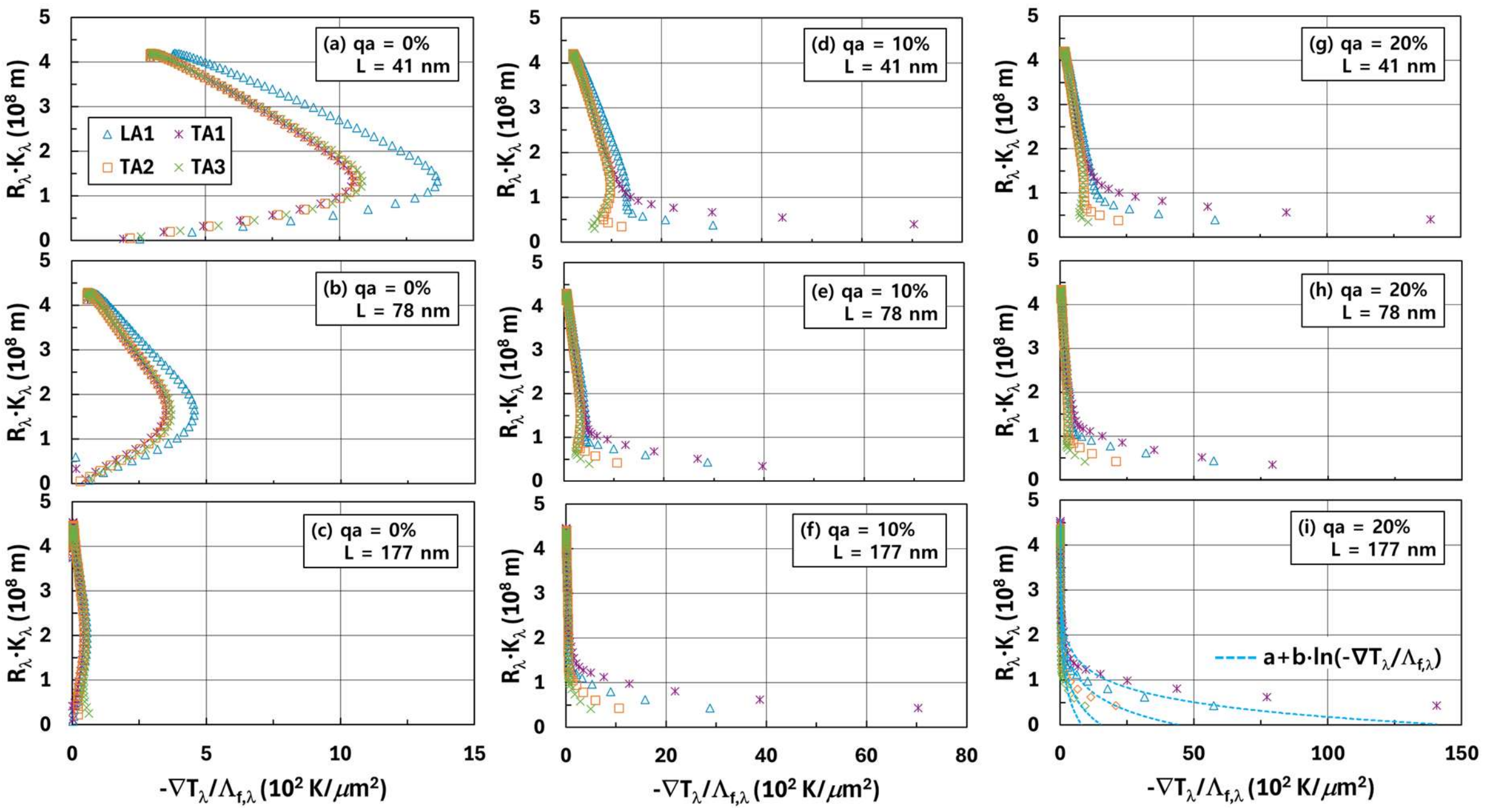

**FIG. 6**. Mode-resolved suppression curves as a function of $\xi_\lambda = -\nabla T_\lambda/\Lambda_{f,\lambda}$ for low-frequency (LF) acoustic modes (LA1, TA1~TA3). The results are shown for all Si layer thicknesses and acoustic-heating fractions. Under diffuse-side conditions ($L$ = 177 nm, $qa$ = 20%), the light blue dashed lines represent the best-fit $a + b \cdot \ln(\xi_\lambda)$ trendlines for each LF mode, demonstrating direct numerical confirmation of the semianalytic $-\ln \xi_\lambda$ scaling. The legend in (a) applies to all panels.

Figures 6 and 7 display the weighted thermal transport resistance (WTRT), $R_\lambda K_\lambda$, as a function of the nonlocal driving parameter $-\nabla T_\lambda/\Lambda_{f,\lambda}$ for the LF (Fig. 6) and HF (Fig. 7) modal groups, respectively, over all combinations of the acoustic heat fraction $qa$ and Si layer thickness $L$. Each symbol corresponds to one of the FPDM acoustic bands (LA1–LA6, TA1–TA6), which are plotted separately but evaluated under the same nanoscale hotspot conditions. To assess whether the modal WTRT follows a logarithmic dependence, we fit the numerical data to the form

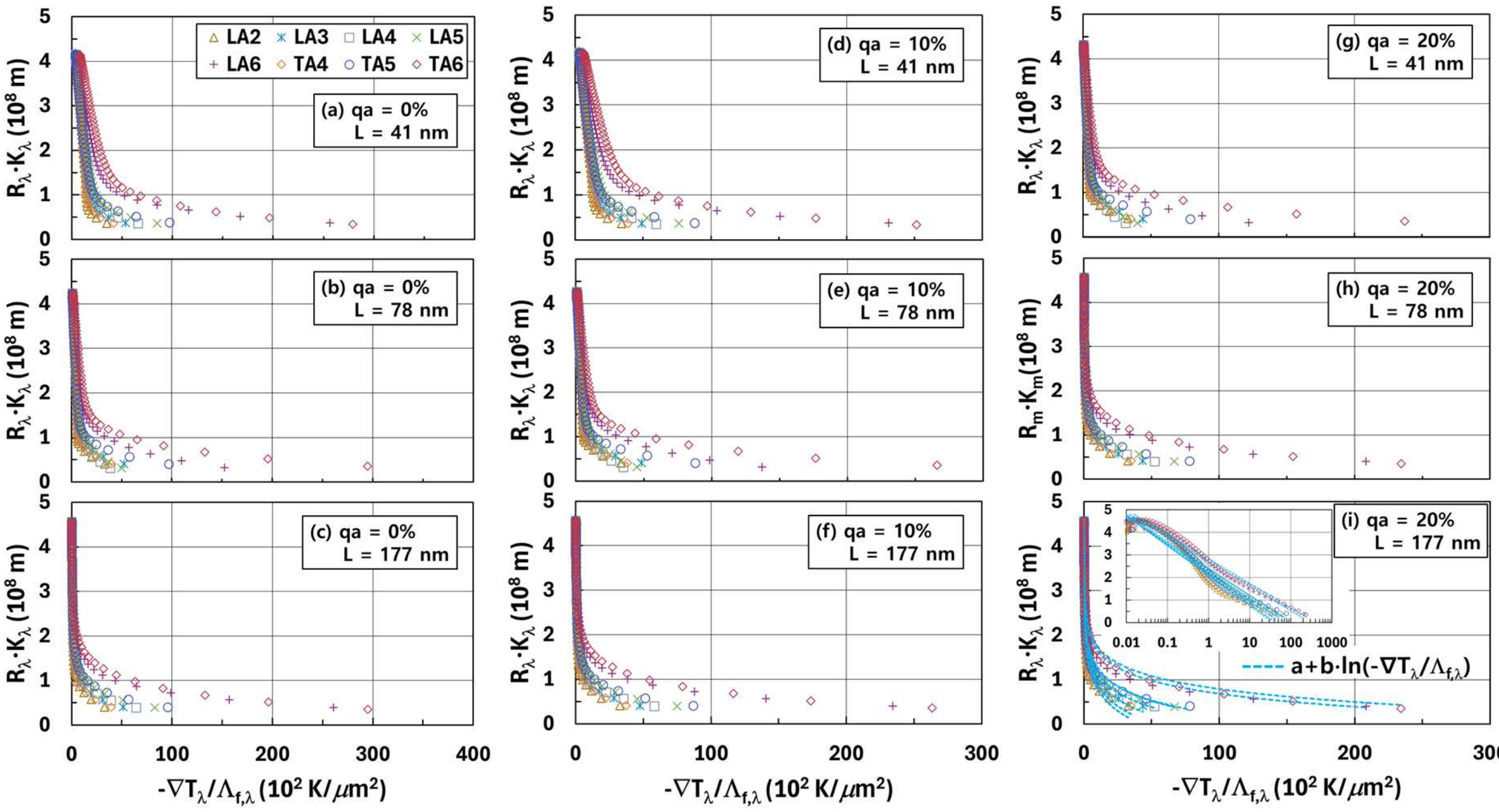

**FIG. 7**. Mode-resolved suppression curves as a function of $\xi_\lambda = -\nabla T_\lambda/\Lambda_{f,\lambda}$ for high-frequency (HF) acoustic modes (LA2–LA6, TA4–TA6). The results are shown for all Si layer thicknesses and acoustic-heating fractions. Under diffuse-side conditions (*L* = 177 nm, *qa* = 20%), the light blue dashed lines in (i) represent the best-fit $a + b \cdot \ln(\xi_\lambda)$ trendlines for each HF mode, demonstrating direct numerical confirmation of the predicted microscopic $-\ln \xi_\lambda$ suppression law. Inset in (i): the same data are replotted with a logarithmic *x*-axis to better visualize the ln-fit over multiple decades of $\xi_\lambda$. The log-*x* replot is shown only in (i) because it corresponds to the representative case for assessing the fit, with ln trendlines, while retaining linear–linear axes in the other panels ensures readability and consistency across Fig. 7. For clarity, the inset is restricted to $\xi_\lambda \geq 2 \times 10^{-2}$, while the full $\xi_\lambda$ range is shown in the main panel; the excluded far-field low-drive regime ($\xi_\lambda \to 0$) occurs far from the hotspot as $\nabla T_\lambda \to 0$ near the isothermal boundaries, where the log-*x* representation can become tightly clustered and is not informative for assessing the ln-fit quality. The legend in (a) applies to all panels.

$$R_\lambda K_\lambda \approx a + b\ln\left(-\frac{\nabla T_\lambda}{\Lambda_{f,\lambda}}\right), \qquad (7)$$

**TABLE II**. Log-fit coefficients $a$ and $b$ and the corresponding $R^2_{\text{fit}}$ values for the LF and HF phonon modes as functions of $qa$ and Si layer thickness $L$. For the LF modes LA1, TA1, TA2, and TA3, the ($a$, $b$) pairs are (2.1300, −0.5609), (2.3526, −0.4716), (1.7352, −0.6342), and (1.5036, −0.7183), respectively. For the HF modes LA2, LA3, LA4, LA5, LA6, TA4, TA5, and TA6, the $(a, b)$ pairs are (2.0843, −0.5543), (2.1941, −0.5071), (2.2768, −0.4835), (2.3694, −0.4716), (2.6341, −0.4230), (2.1248, −0.5334), (2.3368, −0.4600), and (2.7292, −0.4206), respectively. Here, $a$ and $b$ are the fitting parameters in Eq. (7).

| Acoustic heating fraction ($qa$) | Phonon mode (LF/HF) | Si-layer thickness ($L$) | Log-fit $R^2_{\text{fit}}$ | Mean log-fit $R^2_{\text{fit}}$ |
|---|---|---|---|---|
| 0% | LF (LA1, TA1~TA3) | 41 nm | 0.32 ~ 0.41 | 0.35 |
| | | 78 nm | 0.14 ~ 0.38 | 0.22 |
| | | 177 nm | 0.19 ~ 0.70 | 0.41 |
| | HF (LA2~LA6, TA4~TA6) | 41 nm | 0.92 ~ 0.96 | 0.95 |
| | | 78 nm | 0.92 ~ 0.97 | 0.94 |
| | | 177 nm | 0.93 ~ 0.94 | 0.93 |
| 10% | LF (LA1, TA1~TA3) | 41 nm | 0.74 ~ 0.93 | 0.86 |
| | | 78 nm | 0.77 ~ 0.93 | 0.86 |
| | | 177 nm | 0.85 ~ 0.91 | 0.88 |
| | HF (LA2~LA6, TA4~TA6) | 41 nm | 0.93 ~ 0.97 | 0.95 |
| | | 78 nm | 0.92 ~ 0.96 | 0.94 |
| | | 177 nm | 0.92 ~ 0.93 | 0.93 |
| 20% | LF (LA1, TA1~TA3) | 41 nm | 0.86 ~ 0.94 | 0.91 |
| | | 78 nm | 0.85 ~ 0.95 | 0.90 |
| | | 177 nm | 0.89 ~ 0.93 | 0.91 |
| | HF (LA2~LA6, TA4~TA6) | 41 nm | 0.94 ~ 0.97 | 0.96 |
| | | 78 nm | 0.92 ~ 0.96 | 0.94 |
| | | 177 nm | 0.91 ~ 0.92 | 0.91 |

as illustrated by the light blue dashed curves in panels (i) of Figs. 6 and 7 for the cases that visually exhibit the clearest ln-scaling. The quality of this log fit is quantified through the coefficient of determination $R^2_{\text{fit}}$, which is compiled in Table II for all $qa$, $L$, and modal groups.

Taken together, these results strongly contrast the LF and HF modes. HF bands display a robust and nearly universal ln-trend across all film thicknesses and heating partitions, whereas LF bands deviate strongly from ln-scaling when acoustic heating is weak and only recover logarithmic behavior under conditions that promote quasiballistic participation of acoustic modes.

### B. LF phonons: Thickness-dependent emergence of log-scaling

In contrast to the HF sector, the LF modes in Fig. 6 display a much richer and strongly thickness-dependent response. For a small acoustic heating fraction ($qa$ = 0%), the LF WTRT curves deviate markedly from a simple logarithmic trend, particularly at larger values of $\xi_\lambda = -\nabla T_\lambda/\Lambda_{f,\lambda}$, and the different spectral bands fail to collapse onto a single universal trajectory. This behavior is quantified by the poor quality of the logarithmic fits at $qa$ = 0%, where $R_{\text{fit}}^2$ ranges only from approximately 0.14 to 0.70 depending on $L$ (Table II). Under these conditions, the LF phonons are weakly excited and reside predominantly in the fully ballistic regime, for which the semianalytical model predicts saturation rather than ln-type accumulation.

As $qa$ is increased to 10% and 20%, the LF modes become more strongly populated and are driven into a regime where their MFPs are comparable to the effective nonlocal length scale set by the Si layer thickness. Consistent with this trend, the corresponding panels in Fig. 6 show that the LF WTRT curves progressively straighten and increasingly align with the logarithmic trend lines, most clearly in Fig. 6(i) for $qa$ = 20% and $L = 177$nm. The associated log-fit quality improves dramatically: the mean $R_{\text{fit}}^2$ rises to approximately 0.86 at $qa$ = 10% and to 0.90~0.91 at $qa$ = 20%. These results indicate that ln-scaling for LF phonons is not an intrinsic property of the LF modes themselves, but instead emerges only when the acoustic modes are sufficiently excited and rendered effectively quasiballistic, rather than strictly ballistic, with respect to the relevant device length scales.

### C. HF phonons: universal log-scaling across all film thicknesses

For the HF modes (Fig. 7), the WTRT curves collapse onto a narrow family of logarithmic-like trajectories for all three thicknesses $L$ = 41, 78, 177 nm and for all values of $qa$. The light blue dashed curves in Fig. 7(i) demonstrate that a simple ln fit of the form $a + b\ln(-\nabla T_\lambda/\Lambda_{f,\lambda})$ captures the HF behavior remarkably well over the range where the nonlocal drive is appreciable. To better assess the fit quality over the wide dynamic range of $\xi_\lambda \equiv -\nabla T_\lambda/\Lambda_{f,\lambda}$, Fig. 7(i) includes a log-$x$ inset. The log-$x$ replot is shown only in panel (i), which corresponds to the representative case where the ln trendlines are displayed and provides the most informative $\xi_\lambda$ span for evaluating the fit, while the other subpanels are kept in linear–linear form for consistency across Fig. 7. For clarity, the inset is restricted to $\xi_\lambda \geq 2 \times 10^{-2}$, whereas the full $\xi_\lambda$ range remains shown in the main panel; in the far-field low-drive limit ($\xi_\lambda \to 0$) encountered as $\nabla T_\lambda \to 0$ near the isothermal boundaries, the log-$x$ mapping can become tightly clustered and is not informative for assessing the ln-fit. This visual impression is confirmed quantitatively by the log-fit statistics in Table II, where the HF bands yield $R^2_{\text{fit}} \approx$ 0.93~0.96 for all combinations of $qa$ and $L$.

The weak sensitivity of the HF ln-scaling to film thickness is consistent with their comparatively shorter MFPs and shorter effective relaxation times (Table I). Within the relaxation-time approximation (RTA), shorter $\tau_{\text{eff},\lambda}$ corresponds to a higher *effective* scattering rate and thus a more rapid relaxation of modal nonequilibrium toward local equilibrium. As a result, varying $L$ over the present range introduces only a modest additional boundary-induced perturbation to the HF modal response compared with the hotspot-driven nonlocal suppression, and the HF sector therefore retains the same ln-type quasiballistic dependence that characterizes bulk Si. In other words, the ln-regime established in bulk theory is essentially inherited intact by the HF modal sector in the confined Si layer.

## D. Influence of film thickness on quasiballistic scaling

The combined LF and HF results highlight the role of the film thickness in shaping quasiballistic scaling in the confined Si layer. For the HF modes, the ln-trend is essentially thickness independent, which is consistent with their shorter MFPs and predominantly local behavior; their effective transport reflects the hotspot geometry rather than the film boundaries. LF modes, however, are acutely sensitive to $L$. At small thicknesses ($L$ = 41 nm), most LF MFPs remain much larger than the film thickness; thus, the modes behave nearly ballistic, and the WTRT deviates from ln scaling, particularly at low $qa$. As the film becomes thicker (78 and 177 nm), an increasing fraction of LF modes have MFPs comparable to $L$, entering the quasiballistic window where logarithmic dependence can develop.

This thickness dependence is captured both by the shape of the curves in Fig. 6 and by the evolution of $R^2_{\mathrm{fit}}$ in Table II. For a fixed $qa$, the LF log-fit quality tends to improve with increasing $L$, especially at $qa$ = 20%, where the thickest film shows the cleanest ln behavior. These trends corroborate the semianalytic picture in which the upper cutoff $\Lambda_c$ of the ln-active modes scales with the geometric length $L$; only when a substantial portion of the LF spectrum satisfies $\Lambda \lesssim \Lambda_c \sim L$ does ln-scaling become apparent.

### E. Comparison with the bulk quasiballistic model of Ref.[6]

Vermeersch and Mingo's bulk analysis[6] established that the effective thermal conductivity associated with small radial heat sources in crystalline Si exhibits a robust intermediate regime, between the ballistic and diffusive limits, in which $K_{\mathrm{eff}}(R) \propto \ln R$. Within that treatment, all contributing phonons are implicitly aggregated into a single effective spectrum, so that the ln-regime appears as a property of the bulk medium under unbounded geometry.

Our mode-resolved results for the confined Si layer reveal both continuity with, and departures from, this bulk behavior. On the one hand, the HF sector in Fig. 7 clearly reproduces a ln-type dependence with high $R^2_{\mathrm{fit}}$, demonstrating that the basic bulk ln-scaling is preserved for modes whose MFPs remain shorter than or comparable to the relevant nonlocal length scales, even in the presence of film boundaries. On the other hand, the LF sector shows that the ln-regime is not universal across all modes: ballistic LF phonons, which play a key role in approaching the ln-regime in bulk size effects, no longer follow ln scaling in thin films unless their excitation and the film thickness are such that they become quasiballistic relative to $L$.

Thus, while the bulk ln-regime provides a useful reference, the present results demonstrate that in confined Si devices ln-type quasiballistic behavior arises selectively at the modal level rather than as a uniform property of the entire phonon spectrum. This distinction is essential for interpreting and modeling device-level thermal behavior using bulk-based quasiballistic theories.

### F. Physical picture of modal logarithmic suppression

In confined Si devices, the ln regime is governed by a spectrally selective set of modes—high-frequency phonons and quasiballistic portions of the low-frequency spectrum for which the effective nonlocal length is comparable to the MFP and the suppression kernel is nearly constant—while fully ballistic low-frequency modes contribute only an almost uniform background and do not control the ln scaling. This mode-resolved interpretation explains the fragile, geometry-dependent nature of the ln regime in thin Si layers and delineates the excitation and thickness ranges over which logarithmic quasiballistic transport is expected in SOI transistor structures.

The logarithmic decay of the modal heat-carrying capability with increasing nonlocality implies that long-MFP phonons are suppressed only gradually: they are not sharply cut off beyond a characteristic length, but retain finite weight over many decades in the nonlocal parameter and remain active heat carriers that remove a logarithmically reduced yet non-negligible fraction of heat from nanoscale hotspots even under strongly nonlocal conditions.

As the characteristic length scale (hotspot size or film thickness) becomes much smaller than their MFPs, their modal weight saturates and ceases to contribute to further logarithmic accumulation, leading to a substantial reduction of the effective film conductivity relative to the bulk value.

Viewed at the modal level, ln-scale behavior is not merely an ensemble-averaged property of $K(L_c)$ or $K(k)$ but a microscopic manifestation of the nonlocal suppression kernel acting on each phonon mode, as shrinking hotspot sizes in the SOI device layer cause individual modes to sample increasingly nonuniform temperature fields and experience ln-like reductions in effective heat flux. This phonon-modal logarithmic suppression directly connects nonlocal transport physics to device-level thermal management and provides a spectral framework for designing materials and geometries that control the impact of long-MFP phonons in advanced SOI transistors.

## VI. CONCLUSION

We analyzed mode-resolved logarithmic quasiballistic heat transport in thin silicon layers (thicknesses of 41, 78, and 177 nm) representative of SOI transistor structures. A semianalytical framework formulated in the spectral phonon mean-free-path (MFP) domain and benchmarked against a full-phonon-dispersion Boltzmann transport equation (BTE) model for a $10 \times 10$ nm$^2$ hotspot was used to identify the phonon modes that sustain ln-type nonlocal behavior. We found that logarithmic accumulation arises only for modes whose MFPs lie on a log-uniform conductivity plateau and are diffusive-side or quasiballistic with respect to the relevant length scale, whereas fully ballistic long-MFP modes contribute a saturated, nonlogarithmic background. Consequently, the modal heat-carrying capability decays only logarithmically with increasing nonlocality, so long-MFP phonons are suppressed very slowly and remain active heat carriers over many decades in the modal nonlocality parameter. By weighting each mode with its thermal transport resistance, we define a participation metric that reveals distinct behaviors for high- and low-frequency modes. High-frequency modes exhibit a robust logarithmic dependence that is largely insensitive to film thickness. In contrast, low-frequency modes display a thickness- and excitation-dependent crossover from ballistic saturation to logarithmic behavior as their mean free paths become comparable to the layer thickness, a trend that is not captured by the bulk analysis of Vermeersch and Mingo.[6] Taken together, these results show that nonlocality in confined Si layers does not emerge as a collective property of the mode ensemble, but is instead an intrinsic feature of each individual mode.

## SUPPLEMENTARY MATERIAL

See the supplementary material for details of the semianalytic modal framework that explains the thermal-conductivity-weighted logarithmic accumulation across diffusive, quasiballistic, and fully ballistic regimes and its connection to the continuum-scale nonlocal conductivity (Sec. S1), as well as the formulation of the mode-dependent suppression function $\phi_\lambda$ and the resulting thin-film phonon MFPs $\Lambda_{f,\lambda}$ used in this study (Sec. S2).

**Acknowledgments**

We thank the two anonymous reviewers for their careful reading of the manuscript and for their constructive comments, which significantly improved the clarity and quality of this work.

**Supplementary Material for "Mode-resolved logarithmic quasiballistic heat transport in thin silicon layers: Semianalytic and Boltzmann transport analysis"**

**Authors:** Jae Sik Jin*

**Affiliations:** Department of Smart Manufacturing Systems, Chosun College of Science & Technology, Gwangju 61453, Republic of Korea

***Correspondence to:** jinjs@cst.ac.kr

## S1. Semianalytic framework for the thermal conductivity (*K*)-weighted modal logarithmic behavior

Here, we consider a thermal conductivity (*K*)–weighted *modal* phonon quantity and develop a semianalytic framework for its logarithmic behavior. The same framework is then applied sequentially to three transport regimes: (i) bulk-like diffusive-side transport (Section S1.1), (ii) quasi-ballistic transport in thin Si layers (Section S1.2), and (iii) fully ballistic long–mean free path (MFP) modes (Section S1.3). Throughout these Sections, $\Lambda$ denotes the phonon MFP, and $\sigma$ denotes the characteristic nonlocal length scale (e.g., the Si layer thickness or the lateral size of the nanoscale hotspot). (iv) Section S1.4 then introduces an aggregated spectral quantity $Q_{\mathrm{tot}}(\sigma)$ and uses it to connect this modal framework to the continuum-scale nonlocal thermal conductivity $K_{\mathrm{eff}}(\sigma)$, thereby identifying the range of nonlocal lengths $\sigma$ over which $K_{\mathrm{eff}}(\sigma)$ exhibits logarithmic behavior and explaining why this log-type dependence disappears in the fully ballistic and fully diffusive limits.

Importantly, the present semianalytic framework is formulated directly in the spectral MFP domain. Rather than introducing any single 'effective' phonon mode or mode-averaged propagator,[1-6] we explicitly retain the full modal dependence of the thermal conductivity spectrum $K(\Lambda)$, the modal weight $w(\Lambda)$, and the nonlocal suppression kernel $S(\Lambda/\sigma)$. All logarithmic trends discussed below therefore originate from the modal accumulation over the spectral MFP distribution, not from a continuum-level fit to an averaged kernel.

### S1.1 Bulk-like diffusive-side regime ($\Lambda \ll \sigma$, characteristic nonlocal length)

We begin by introducing the spectral thermal conductivity density $K(\Lambda)$. The total lattice thermal conductivity ($K_{\mathrm{tot}}$)is written as

$$K_{\mathrm{tot}} = \int_{\Lambda_{\min}}^{\Lambda_{\max}} K(\Lambda)\, d\Lambda. \qquad \text{(S1)}$$

First-principles calculations and experiments have shown that, over a wide MFP interval in crystalline Si, the conductivity spectrum is nearly uniform per decade in $\Lambda$. This is often expressed as

$$\frac{dK}{d(\ln \Lambda)} \approx K_0. \qquad \text{(S2)}$$

Equation (S2) means that the thermal conductivity spectrum is nearly constant per logarithmic MFP interval (a log-uniform plateau) in the MFP range $\Lambda_{\text{min}} \lesssim \Lambda \lesssim \Lambda_{\text{max}}$. This behavior is equivalent to the approximate form

$$K(\Lambda) \approx \frac{K_0}{\Lambda}, \qquad \text{(S3)}$$

i.e., each logarithmic MFP decade contributes roughly the same amount to $K_{\text{tot}}$. Here, $K_0$ denotes the approximately constant value of the thermal conductivity spectrum per logarithmic MFP interval, i.e., the plateau of $dK/d(\ln \Lambda)$ in Eq. (S2).[7-11]

Near a nanoscale hotspot, the $K$-weighted modal quantity is constructed from three ingredients: the spectral thermal conductivity $K(\Lambda)$; a modal transport weight $w(\Lambda)$, which groups together the modal transport efficiency and the effective driving force (e.g., the local modal temperature drop); and a nonlocal suppression kernel $S(\Lambda/\sigma)$. We therefore write the $K$-weighted modal quantity accumulated up to a cutoff MFP $\Lambda_c$ as

$$Q(\Lambda_c) = \int_{\Lambda_{\text{min}}}^{\Lambda_c} K(\Lambda)\, w(\Lambda)\, S\left(\frac{\Lambda}{\sigma}\right) d\Lambda. \qquad \text{(S4)}$$

In this Section, the independent variable of interest is the cutoff mean free path $\Lambda_c$ at a fixed nonlocal length scale $\sigma$. Throughout Sections S1.1–S1.3, $\sigma$ is held fixed and the $\sigma$-dependence of $Q$ is implicit. We therefore examine how the $K$-weighted modal accumulation $Q(\Lambda_c)$ grows as modes with increasingly long-MFPs are included for a given geometry (fixed $\sigma$). We show that it exhibits a ln $\Lambda_c$ dependence over the diffusive and quasi-ballistic portions of the spectrum. This analysis is purely modal at fixed $\sigma$ and should be distinguished from the continuum-scale question of how the effective nonlocal thermal conductivity $K_{\mathrm{eff}}(\sigma)$ itself varies with the nonlocal length; that dependence is obtained in Section S1.4 by integrating over the full spectrum and treating $\sigma$ as the independent variable in the aggregated quantity $Q_{\mathrm{tot}}(\sigma)$.

In the bulk-like diffusive-side regime, where $\Lambda$ is much smaller than the nonlocal length scale $\sigma$, the following approximations hold:

- The modal weight varies weakly with $\Lambda$:

$$w(\Lambda) \approx w_0, \qquad \text{(S5)}$$

  with $w_0$ a mode- and geometry-dependent constant.

- The suppression kernel is close to unity and nearly independent of $\Lambda$:

$$S\left(\frac{\Lambda}{\sigma}\right) \approx S_{\mathrm{diff}}, \qquad \text{(S6)}$$

  where $S_{\mathrm{diff}}$ is an order-unity constant.

Using Eqs. (S3)–(S6), the integrand of Eq. (S4) in the diffusive-side interval becomes

$$K(\Lambda)\, w(\Lambda)\, S\left(\frac{\Lambda}{\sigma}\right) \approx \frac{K_0}{\Lambda}\, w_0\, S_{\mathrm{diff}} \;\propto\; \frac{1}{\Lambda}. \qquad \text{(S7)}$$

Hence, in the bulk-like diffusive-side regime the $K$-weighted accumulation up to $\Lambda_c$ is

$$Q_{\mathrm{diff}}(\Lambda_c) \propto \int_{\Lambda_{\min}}^{\Lambda_c} \frac{d\Lambda}{\Lambda} = \ln\left(\frac{\Lambda_c}{\Lambda_{\min}}\right). \qquad \text{(S8)}$$

It is often convenient to introduce the logarithmic modal coordinate

$$\xi \equiv \ln\left(\frac{\Lambda}{\Lambda_0}\right), \qquad \text{(S9)}$$

where $\Lambda_0$ is a reference MFP. In terms of $\xi$, Eq. (S8) can be written as

$$Q_{\mathrm{diff}}(\xi_c) \propto \xi_c - \xi_{\min}, \qquad \text{(S10)}$$

i.e., a linear dependence in the log-MFP variable $\xi$ is equivalent to a logarithmic dependence in the MFP variable $\Lambda$. In the present bulk-like diffusive regime, this implies that the $K$-weighted modal accumulation $Q_{\mathrm{diff}}(\Lambda_c)$ scales as $\ln \Lambda_c$. Here "bulk-like diffusive-side" refers only to phonon modes with $\Lambda \ll \sigma$ in MFP space at a fixed nonlocal length $\sigma$, for which the kernel $S(\Lambda/\sigma)$ is essentially constant; it does not mean that the macroscopic nonlocal thermal conductivity $K_{\mathrm{eff}}(\sigma)$ in the bulk limit itself scales as $\ln \sigma$. The latter continuum-scale behavior is controlled by how the aggregated spectral quantity $Q_{\mathrm{tot}}(\sigma)$, introduced in Section S1.4, changes as the boundary $\Lambda \sim \sigma$ between diffusive-side and ballistic modes moves across the log-uniform plateau of $K(\Lambda)$.

**S1.2 Quasi-ballistic regime in thin Si layer ($\Lambda \sim L$)**

We now extend the above framework to quasi-ballistic low-frequency (LF) modes in a thin Si layer. Consider phonon modes whose MFP is comparable to the Si layer thickness $L$ ($\Lambda \sim L$). In this case, the characteristic nonlocal length $\sigma$ in Eq. (S4) is set by $L$. For quasi-ballistic modes with $\Lambda$ of order $L$ but not much larger, the suppression kernel remains of order unity and varies only weakly with $\Lambda$:

$$S\left(\frac{\Lambda}{L}\right) \approx S_{\mathrm{QB}}, \qquad \text{(S11)}$$

where $S_{\mathrm{QB}}$ is a quasi-ballistic suppression constant. This is the natural extension of Eq. (S6) from the strictly diffusive-side regime ($\Lambda \ll \sigma$) to the diffusive-to-quasi-ballistic interval, i.e.,

$$\Lambda_{\mathrm{min}} \lesssim \Lambda \lesssim \Lambda_{\mathrm{diff/QB}} \sim L. \qquad \text{(S12)}$$

Here, $\Lambda_{\mathrm{diff/QB}}$ denotes the characteristic MFP at the diffusive-to–quasi-ballistic crossover, i.e., the upper end of the diffusive-side regime, which is of the order of the Si layer thickness $L$ in the present geometry. In this interval, the log-uniform conductivity spectrum [Eq. (S2)] remains valid, and the modal weight $w(\Lambda)$ is only weakly dependent on $\Lambda$ and can therefore be approximated by $w_0$. Moreover, the kernel $S(\Lambda/L) \approx S_{\mathrm{QB}}$ does not introduce any additional $\Lambda$ dependence. Substituting these approximations into Eq. (S4) yields the quasi-ballistic accumulation

$$Q_{\mathrm{QB}}(\Lambda_c) \propto \int_{\Lambda_{\mathrm{min}}}^{\Lambda_c} \frac{d\Lambda}{\Lambda} = \ln\left(\frac{\Lambda_c}{\Lambda_{\mathrm{min}}}\right) \qquad \text{(S13)}$$

for $\Lambda_c \leq \Lambda_{\mathrm{diff/QB}} \approx L$. Thus, quasi-ballistic LF modes with $\Lambda$ comparable to $L$ preserve the $1/\Lambda$ integrand structure and generate the *same logarithmic dependence* on $\Lambda$ (or linear dependence on $\xi$) as the bulk-like diffusive-side modes. In other words, quasi-ballistic modes continue the bulk logarithmic trend up to the crossover scale set by $L$. Therefore, in the quasi-ballistic regime the $K$-weighted modal accumulation also scales as $Q_{\mathrm{QB}}(\Lambda) \propto \ln \Lambda$.

### S1.3 Fully ballistic long-MFP modes ($\Lambda \gg L$)

Finally, we consider fully ballistic LF modes with MFPs much longer than the Si layer thickness $L$, i.e., $\Lambda \gg L$. In this limit, heat transport across the Si layer becomes purely ballistic. The streaming part of the FPDM–BTE then reduces to the standard gray, single-MFP problem. The well-known analytic solution of this gray problem provides the asymptotic form of the suppression kernel:

$$S_{\mathrm{B}}\left(\frac{\Lambda}{L_{\mathrm{Si}}}\right) \propto \frac{L}{\Lambda} \qquad \text{(S14)}$$

for $\Lambda \gg L$. (Here, the proportionality constant is an order-unity factor determined by boundary conditions and is not important for the logarithmic versus saturated behavior.)

Substituting the ballistic kernel (S14) into the general modal expression (S4), and using the log-uniform spectrum (S3) and weak Λ-dependence of $w(\Lambda)$, we obtain the asymptotic ballistic integrand

$$K(\Lambda)\, w(\Lambda)\, S_{\mathrm{B}}\left(\frac{\Lambda}{L}\right) \propto \frac{1}{\Lambda}\,\frac{L}{\Lambda} = \frac{L}{\Lambda^2}. \qquad \text{(S15)}$$

Therefore, the contribution of fully ballistic LF modes to the *K*-weighted quantity scales as

$$Q_{\mathrm{B}}(\Lambda_{\mathrm{max}}) \propto \int_{\Lambda_0}^{\Lambda_{\mathrm{max}}} \frac{L}{\Lambda^2}\, d\Lambda = L\left(\frac{1}{\Lambda_0} - \frac{1}{\Lambda_{\mathrm{max}}}\right), \qquad \text{(S16)}$$

where $\Lambda_0$ is a lower bound of the ballistic interval ($\Lambda_0 \gg L$). As $\Lambda_{\max}$ becomes very large ($\Lambda_{\max} \gg \Lambda_0$), the second term in parentheses becomes negligible, and the ballistic contribution approaches a finite constant:

$$Q_{\mathrm{B}}(\Lambda_{\mathrm{max}} \to \infty) \propto \frac{L}{\Lambda_0} = \text{constant}. \qquad \text{(S17)}$$

Thus, the fully ballistic contribution saturates to a constant and does not contain any term of the form ln Λ. In summary, the *K*-weighted logarithmic decay arises from modes with $\Lambda \lesssim L$, whereas fully ballistic modes with $\Lambda \gg L$ provide only a saturated background without any logarithmic contribution.

This separation directly explains why the log-type nonlocal dependence $K_{\mathrm{eff}}(\sigma) \propto \ln \sigma$ discussed in Section S1.4 is restricted to the quasi-ballistic range where the nonlocal length $\sigma$ lies within the log-uniform plateau $[\Lambda_1, \Lambda_2]$ of the conductivity spectrum, and why both the fully ballistic limit ($\sigma \ll \Lambda_1$) and the fully diffusive bulk limit ($\sigma \gg \Lambda_2$) are free of any $\ln \sigma$ term.

### S1.4 From modal accumulation $Q(\Lambda)$ to continuum-scale $K_{\mathrm{eff}}(\sigma)$

To connect this modal picture in Λ-space to the continuum-scale nonlocal thermal conductivity $K_{\mathrm{eff}}(\sigma)$, we now treat the nonlocal length $\sigma$ itself as the independent variable and integrate over the full log-uniform plateau of the conductivity spectrum. This step identifies the range of $\sigma$ over which the modal logarithmic accumulation discussed in Sections S1.1–S1.3 translates into a log-type nonlocality $K_{\mathrm{eff}}(\sigma) \propto \ln \sigma$, and shows explicitly why this logarithmic behavior disappears again in the fully ballistic ($\sigma \ll \Lambda_1$) and fully diffusive bulk ($\sigma \gg \Lambda_2$) limits. To this end we introduce a simplified aggregated spectral quantity $Q_{\mathrm{tot}}(\sigma) = \int_0^{\infty} K(\Lambda)\, S(\Lambda/\sigma)\, \mathrm{d}\Lambda$, where $\sigma$ denotes the relevant nonlocal length (e.g., film thickness, hotspot radius, or thermal-grating period). For crystalline Si, first-principles calculations show that the conductivity spectrum is approximately log-uniform over a broad MFP range $[\Lambda_1, \Lambda_2]$, i.e., $K(\Lambda) \simeq \frac{K_0}{\Lambda}$ for $\Lambda_1 \lesssim \Lambda \lesssim \Lambda_2$. In the same plateau, the FPDM–BTE suppression kernel reduces to two simple asymptotic forms: $S\left(\frac{\Lambda}{\sigma}\right) \approx S_0\ (\Lambda \ll \sigma)$ and $S\left(\frac{\Lambda}{\sigma}\right) \approx C_b \frac{\sigma}{\Lambda}\ (\Lambda \gg \sigma)$, where $S_0$ and $C_b$ are order-unity constants that depend only weakly on geometry.

When the nonlocal length lies inside the plateau, $\Lambda_1 \ll \sigma \ll \Lambda_2$, the integral can be split into a "diffusive-side + quasi-ballistic" part ($\Lambda < \sigma$) and a fully ballistic part ($\Lambda > \sigma$):

$$Q_{\mathrm{tot}}(\sigma) \simeq \underbrace{\int_{\Lambda_1}^{\sigma} \frac{K_0}{\Lambda} S_0\, \mathrm{d}\Lambda}_{\text{diffusive-side + quasi-ballistic modes}} + \underbrace{\int_{\sigma}^{\Lambda_2} \frac{K_0}{\Lambda} C_b \frac{\sigma}{\Lambda}\, \mathrm{d}\Lambda}_{\text{fully ballistic modes}}. \qquad \text{(S18)}$$

The first contribution evaluates to $Q_{\Lambda<\sigma}(\sigma) \simeq K_0 S_0 \ln(\frac{\sigma}{\Lambda_1})$, which contains an explicit ln $\sigma$ dependence. By contrast, the ballistic contribution approaches a constant as $\Lambda_2 \to \infty$: $Q_{\Lambda>\sigma}(\sigma) \simeq K_0 C_b (1 - \frac{\sigma}{\Lambda_2}) \approx$ const. Thus, in the intermediate (quasi-ballistic) range $\Lambda_1 \ll \sigma \ll \Lambda_2$ one obtains the simple scaling

$$Q_{\mathrm{tot}}(\sigma) \propto \ln \sigma + (\mathrm{constant}), \qquad \text{(S19)}$$

which directly yields the well-known log-type nonlocal behavior $K_{\mathrm{eff}}(\sigma) \sim \ln \sigma$. In other words, the continuum-scale logarithmic nonlocality arises from the modal logarithmic accumulation of diffusive-side and quasi-ballistic modes with $\Lambda \lesssim \sigma$, whereas fully ballistic long-MFP modes with $\Lambda \gg \sigma$ contribute only a saturated background without any logarithmic term.

In the fully diffusive bulk limit where the nonlocal length $\sigma$ is much larger than the upper end of the log-plateau ($\sigma \gg \Lambda_2$), all contributing modes satisfy $\Lambda \ll \sigma$. In this regime the FPDM–BTE suppression kernel reduces to its diffusive asymptote, $S(\Lambda/\sigma) \approx S_{\mathrm{diff}} \approx$ const., so the spectral integral for $Q_{\mathrm{tot}}(\sigma)$ [and thus for $K_{\mathrm{eff}}(\sigma)$] becomes independent of $\sigma$. As a result, $K_{\mathrm{eff}}(\sigma)$ simply saturates to a constant and any ln $\sigma$–type scaling disappears. The logarithmic nonlocality $K_{\mathrm{eff}}(\sigma) \propto \ln \sigma$ is therefore confined to the intermediate quasi-ballistic range $\Lambda_1 \ll \sigma \ll \Lambda_2$ and is absent in both the fully ballistic and fully diffusive bulk limits. In particular, the diffusive-side approximation $S(\Lambda/\sigma) \approx$ const. used in Section S1.1 explains the logarithmic accumulation of $Q(\Lambda_c)$ with respect to $\Lambda_c$ at fixed $\sigma$, but does not, by itself, imply a logarithmic dependence of the bulk nonlocal conductivity on $\sigma$. The ln $\sigma$ dependence of $K_{\mathrm{eff}}(\sigma)$ arises only while the moving boundary $\Lambda \sim \sigma$ between diffusive-side and ballistic modes sweeps across the log-uniform plateau of $K(\Lambda)$; once $\sigma$ lies well beyond this plateau in the fully diffusive bulk regime, the accumulated spectrum is saturated and the ln $\sigma$ term vanishes.

**S2. Calculation of the suppression function $\phi_\lambda$**

In the main text, the phonon MFP in the thin film, $\Lambda_{f,\lambda}$, is estimated from the bulk MFP $\Lambda_{b,\lambda}$ via

$$\Lambda_{f,\lambda} = \phi_\lambda \Lambda_{b,\lambda}, \qquad \text{(S20)}$$

where $\phi_\lambda$ is a suppression function that accounts for additional boundary scattering in the thin film. Following the approach of Ref.[19] for in-plane thermal transport in silicon thin films, $\phi_\lambda$ is given by

$$\phi_\lambda = 1 - \frac{1-p}{\delta} \frac{1-\exp(-\delta)}{1-p\ \exp(-\delta)}, \qquad \text{(S21)}$$

where $\delta = L/\Lambda_{b,\lambda}$ is a dimensionless parameter defined as the ratio of the film thickness $L$ to the bulk phonon mean free path $\Lambda_{b,\lambda}$, and $p$ is the specularity parameter of the film boundaries.

For each phonon band $\lambda$, we first evaluate $\delta$ from the corresponding bulk mean free path $\Lambda_{b,\lambda}$ and the film thickness $L$ considered in this study. The suppression function $\phi_\lambda$ is then computed using Eq. (S21). The resulting values of $\phi_\lambda$ for all bands are summarized in Fig. S1. As expected, $\phi_\lambda$ decreases for modes with longer bulk mean free paths (stronger impact of boundary scattering), whereas $\phi_\lambda \to 1$ in the diffusive limit $\delta \gg 1$ or for nearly specular boundaries $(p \to 1)$. These band-resolved suppression factors are subsequently used to determine the thin-film phonon mean free paths $\Lambda_{f,\lambda} = \phi_\lambda \Lambda_{b,\lambda}$ in the BTE calculations presented in the main text.

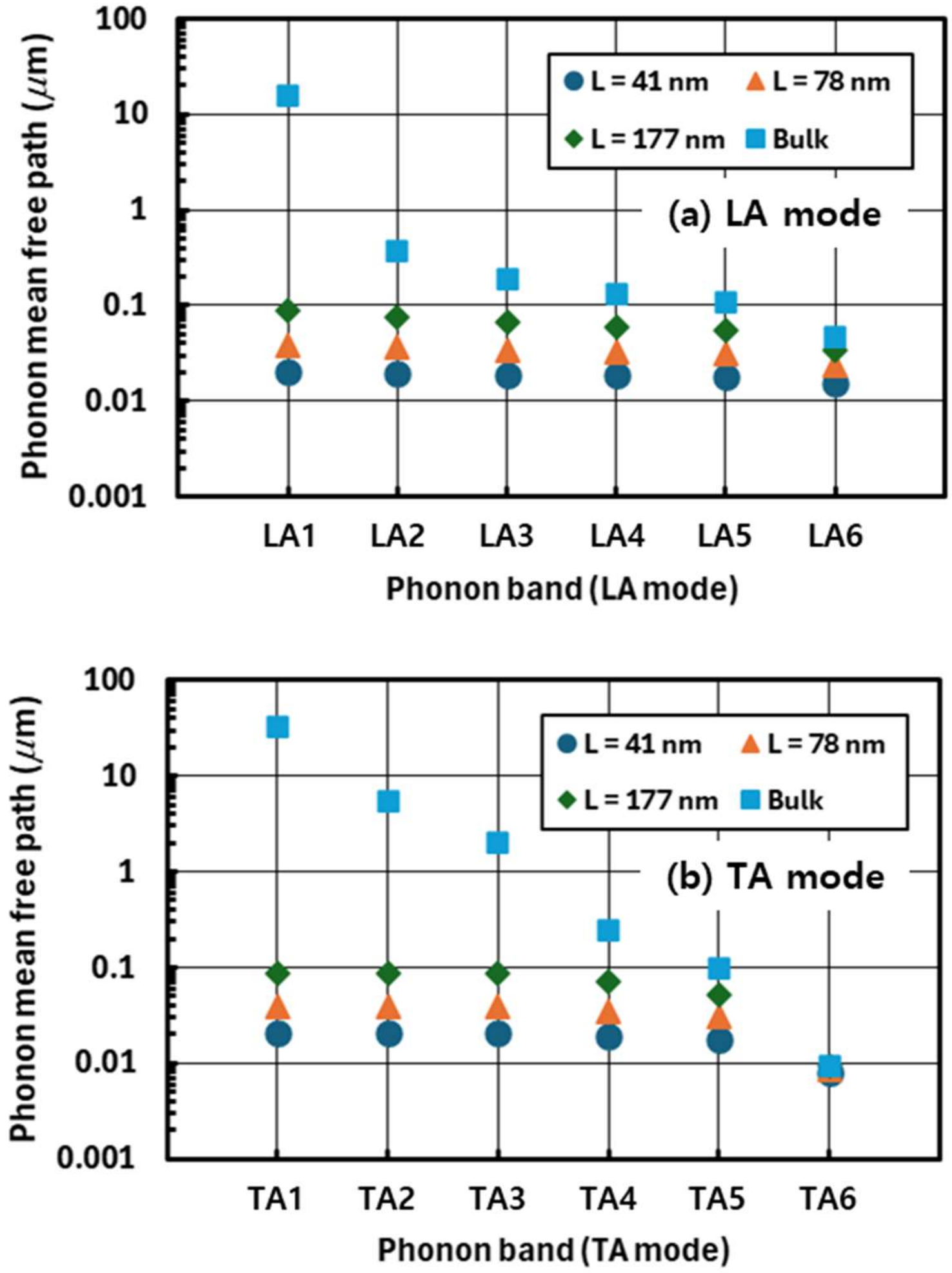

**FIG. S1**. Suppression function $\phi_\lambda$ for (a) LA and (b) TA modes in the silicon thin film, calculated from Eq. (S21). These values are used to obtain the thin-film phonon MFPs $\Lambda_{f,\lambda}$ employed in the phonon BTE calculations.